\documentclass{ws-procs9x6.25Dec2007}

\def\today{28~December~2007}

\def\FWM{FWM}
\def \D {\hbox{d}}
\def \barA {\overline{A}}
\def \barC {\overline{C}}
\def \barD {\overline{D}}

\def \de  {\Delta     \varepsilon}
\def \dc  {\Delta \bar\varepsilon}

\def \mod#1{\vert #1 \vert}
\def \sech{\mathop{\rm sech}\nolimits}

\begin{document}

\title{
Analytic structure of the four-wave mixing model
in photorefractive materials
\footnote{Waves and stability in continuous media,
30 June-7 July 2007, Scicli (Rg), Italy.}
}

\author{Robert Conte}

\address{Service de physique de l'\'etat condens\'e (CNRS URA 2464)
\\ CEA--Saclay, F--91191 Gif-sur-Yvette Cedex, France
\\ E-mail:  Robert.Conte@cea.fr
}

\author{Svetlana Bugaychuk}

\address{Institute of Physics of the National Academy of Sciences of Ukraine
\\ 46 Prospect Nauki, Kiev-39,  UA 03039, Ukraine
\\ E-mail:  bugaich@iop.kiev.ua
}

\centerline{\today}

\begin{abstract}
In order to later find explicit analytic solutions,
we investigate the singularity structure of a fundamental model
of nonlinear optics, the four-wave mixing model
in one space variable $z$.
This structure is quite similar, and this is not a surprise,
to that of the cubic complex Ginzburg-Landau equation.
The main result is that, in order to be single valued,
time-dependent solutions should depend on the space-time coordinates
through the reduced variable
$\xi=\sqrt{z} e^{-t / \tau}$,
in which $\tau$ is the relaxation time.
\end{abstract}


\bigskip\smallskip

\noindent
\textit{PACS 2001} 
  02.30.Jr,
  42.65    



\bodymatter

\section{The four-wave mixing model}

Dynamic holography relies on the ability of the refractive index of a nonlinear media to locally change under the action of light,
i.e.~under the so-called photorefractive response.
The most conventional scheme for dynamic holographic applications consists
in the
formation of a fringe interference pattern between the interacting waves (the light grating),
which creates the refractive index grating.
During the wave interaction, these are the same waves
which create the grating pattern and diffract on this grating.
There exist many applications based on
dynamic holography with photorefractive crystals \cite{BRMK}.

In the four-wave mixing model,
the grating, i.e. the modulation of the refractive index induced by the light interference pattern,
is created by two pairs of co-propagating waves,
1--2 and 3--4.
Let us denote
$A_1,A_2,A_3,A_4$ the complex amplitudes of the four waves,
$\de$ the complex amplitude of the index grating,
$\tau$ the characteristic time of the evolution,
$I_0$ the total intensity of light (a real positive constant),
$\gamma$ the photorefractive coupling (a complex constant).
The four-wave mixing model is the system of
five complex partial differential equations
obeyed by these five complex amplitudes~\cite{BKMPR}
(bar denotes complex conjugation)
\begin{eqnarray}
& &
\left\lbrace
\begin{array}{ll}
\displaystyle{
\partial_z     A_1=-i \de     A_2,\
\partial_z \barA_2= i \de \barA_1,\
\partial_z \barA_3=-i \de \barA_4,\
\partial_z     A_4= i \de     A_3,\
}
\\
\displaystyle{
\left(\partial_t +\frac{1}{\tau} \right) \de
=\frac{\gamma}{I_0} \left(A_1 \barA_2 + \barA_3 A_4 \right).
}
\end{array}
\right.
\label{eq4wm}
\end{eqnarray}
The correspondence of notation with Ref.~\cite{BKK} is
\begin{eqnarray}
& &
 (A_1,A_2,A_3,A_4,Q,\gamma) \hbox{ of Ref.~}
=
(A_1,\barA_4,\barA_3,A_2,-i \de,i \gamma)_{\rm here}.
\end{eqnarray}

At present time, no analytic solution is known to this system,
except a stationary solution in which the grating amplitude
has a pulse profile~\cite{BRMK},
\begin{eqnarray}
& &
\de=K \sech k (z-z_0),\ (K,k)=\hbox{real constants}.
\end{eqnarray}
The purpose of this short article is to investigate the structure of the
singularities of the time-dependent solutions ($\partial_t \not=0$)
in order to find, in a forthcoming paper,
closed form analytic solutions
by making suitable assumptions dictated by the singularity structure.

The first four equations (\ref{eq4wm}), which do not depend on $\partial_t$,
admit six quadratic first integrals.
Two of them,
\begin{eqnarray}
& &
\mod{A_1}^2 + \mod{A_2}^2 = f_1^2(t),\
\mod{A_3}^2 + \mod{A_4}^2 = f_2^2(t),
\label{eq2FI}
\end{eqnarray}
allow us to represent the four complex amplitudes $A_j(z,t)$
with two real functions $f_1,f_2$ of $t$ and six real functions
$\theta_1,\theta_2,
\varphi_1,\varphi_2,\varphi_3,\varphi_4$ of $(z,t)$,
\begin{eqnarray}
& & {\hskip -8.0truemm}
A_1=f_1 \sin \theta_1 e^{i \varphi_1},
A_2=f_1 \cos \theta_1 e^{i \varphi_2},
A_3=f_2 \cos \theta_2 e^{i \varphi_3},
A_4=f_2 \sin \theta_2 e^{i \varphi_4}.
\label{eqAj}
\end{eqnarray}
The four other quadratic first integrals
are pairwise complex conjugate
\begin{eqnarray}
& &
\left\lbrace
\begin{array}{ll}
\displaystyle{
A_1 A_3 + A_2 A_4 = f_3(t),\
\barA_1 \barA_3 + \barA_2 \barA_4 = \overline{f_3(t)},
}
\\
\displaystyle{
\barA_1 A_4 - \barA_2 A_3 = f_4(t),\
A_1 \barA_4 - A_2 \barA_3 = \overline{f_4(t)},\
}
\end{array}
\right.
\label{eq4FI}
\end{eqnarray}
and constrained by one relation,
\begin{eqnarray}
& &
-f_1^2 f_2^2 + \mod{f_3}^2 + \mod{f_4}^2=0.
\label{eqFIrelation}
\end{eqnarray}

\section{Local singularity analysis of the \FWM\ model}
\label{sectionLocal}

The dependent variables $A_j,\de$ generically present movable singularities,
i.e.~singularities whose location in the complex domain
depends on the initial conditions.
The study of the behaviour of $A_j,\de$ near these movable singularities
is a prerequisite to the possible obtention of closed form analytic
solutions to the nonlinear system (\ref{eq4wm}),
so let us first perform it.
Such a study is made by performing the successive steps
of the so-called \textit{Painlev\'e test},
a procedure explained in detail
in Ref.~\cite{Cargese1996Conte,CetraroConte}.

The movable singularities of a PDE
lay on a manifold represented by the equation \cite{WTC}
\begin{eqnarray}
& &
\varphi(z,t) - \varphi_0=0,
\label{eqPDEManifold}
\end{eqnarray}
in which $\varphi$ is an arbitrary function of the independent variables,
and
$\varphi_0$ an arbitrary movable constant.
In order to express the behaviour of the
ten dependent variables $A_j,\barA_j,\de,\dc$,
it is convenient to
introduce an expansion variable $\chi(z,t)$ \cite{Cargese1996Conte}
which vanishes as $\varphi-\varphi_0$
and which is defined by its gradient
\begin{eqnarray}
\chi_z
& = &
 1 + \frac{S}{2} \chi^2,\
\chi_t
 =
 - C + C_z \chi  - \frac{1}{2} (C S + C_{zz}) \chi^2.
\label{eqChixt}
\end{eqnarray}
with the cross-derivative condition
\begin{eqnarray}
X & \equiv &
 S_t + C_{zzz} + 2 C_z S + C S_z = 0.
\label{eqCrossXT}
\end{eqnarray}

Since they will be needed later,
we give the dependence of $\chi,S,C$ on $\varphi$,
\begin{eqnarray}
& &
 \chi
=\frac{\varphi-\varphi_0}{\varphi_z
- \displaystyle{\frac{\varphi_{zz}}{2 \varphi_z}}
(\varphi-\varphi_0)}
=\left[\frac{\varphi_z}{\varphi-\varphi_0}
- \frac{\varphi_{zz}}{2 \varphi_z}\right]^{-1},\
\varphi_z \not=0,
\label{eqchi}
\\
& &
S=\lbrace \varphi;z \rbrace
=\frac{\varphi_{zzz}}{\varphi_z}
 - \frac{3}{2} \left(\frac{\varphi_{zz}} {\varphi_z} \right)^2,\
 C=- \varphi_t / \varphi_z.
\label{eqS}
\label{eqC}
\end{eqnarray}

\subsection{Leading behaviour}
\label{sectionLeading}

The first step is to
determine all possible families of noncharacteristic
(i.e.~$\varphi_x \varphi_t \not=0$)
movable singularities,
i.e.~all the leading behaviours
\begin{eqnarray}
& &
\chi \to 0:\
\left\lbrace
\begin{array}{ll}
\displaystyle{
A_k \sim a_k \chi^{p_k},\ \barA_k \sim b_k \chi^{P_k},\ a_k b_k \not=0,\
k=1,2,3,4,\
}
\\
\displaystyle{
\de \sim q_0 \chi^\alpha,\ \dc \sim r_0 \chi^\beta,\ q_0 r_0 \not=0,
}
\end{array}
\right.
\label{eqLeadingBehaviour}
\end{eqnarray}
in which the exponents $p_k,P_k,\alpha,\beta$ are not all positive integers.
If one assumes that
the two terms $A_1 \barA_4$ and $A_3 \barA_2$ have the same singularity order,
the ten exponents $p_k,P_k,\alpha,\beta$ obey the twelve linear equations,
\begin{eqnarray}
& &
\left\lbrace
\begin{array}{ll}
\displaystyle{
p_1-1=p_2+\alpha,\ \ P_2-1=P_1+\alpha,\
}\\ \displaystyle{
p_4-1=p_3+\alpha,\ \ P_3-1=P_4+\alpha,\
}\\ \displaystyle{
P_1-1=P_2+\beta ,\ \ p_2-1=p_1+\beta ,\
}\\ \displaystyle{
P_4-1=P_3+\beta ,\ \ p_3-1=p_4+\beta ,\
}\\ \displaystyle{
\alpha-1=p_1+P_2=p_4+P_3,\
}\\ \displaystyle{
\beta -1=P_1+p_2=P_4+p_3,
}
\end{array}
\right.
\label{eqLeadingPowers}
\end{eqnarray}
and the ten coefficients $a_k,b_k,q_0,r_0$ obey the ten nonlinear
equations
\begin{eqnarray}
& &
\left\lbrace
\begin{array}{ll}
\displaystyle{
p_1 a_1 = - q_0 a_2,\ p_2 a_2 = -r_0 a_1,\ p_4 a_4 = -q_0 a_3,\ p_3 a_3 =-r_0 a_4,\
}
\\
\displaystyle{
P_1 b_1 = r_0 b_2,\ P_2 b_2 =  q_0 b_1,\ P_4 b_4 = r_0 b_3,\ P_3 b_3 = q_0 b_4,\
}
\\
\displaystyle{
C \alpha q_0 =-(     \gamma/I_0) (a_1 b_2 + a_4 b_3),
}
\\
\displaystyle{
C \beta  r_0 =-(\bar \gamma/I_0) (b_1 a_2 + b_4 a_3).
}
\end{array}
\right.
\label{eqLeadingCoeffs}
\end{eqnarray}
Therefore the five squared moduli $\mod{A_k}^2,\mod{\de}^2$
behave like double poles,
\begin{eqnarray}
& &
p_k+P_k=-2,\ k=1,2,3,4,\ \alpha+\beta=-2,\
\label{eqLeadingPowersa}
\end{eqnarray}
and the linear system (\ref{eqLeadingPowers}) is solved as
\begin{eqnarray}
& & {\hskip -9.0truemm}
\left\lbrace
\begin{array}{ll}
\displaystyle{
p_1=-1+s_1+\delta,\
P_1=-1-s_1-\delta,\
}\\ \displaystyle{
p_2=-1+s_1-\delta,\
P_2=-1-s_1+\delta,\
}\\ \displaystyle{
p_4=-1+s_2+\delta,\
P_4=-1-s_2-\delta,\
}\\ \displaystyle{
p_3=-1+s_2-\delta,\
P_3=-1-s_2+\delta,\
}\\ \displaystyle{
\alpha=-1+2 \delta,\
\beta =-1-2 \delta,\
}
\end{array}
\right.
\label{eqLeadingPowersSol}
\end{eqnarray}
in which $\delta,s_1,s_2$ are to be determined by the nonlinear system
(\ref{eqLeadingCoeffs}).

The first set of relations
\begin{eqnarray}
& &
-q_0 r_0=p_1 p_2=p_4 p_3=P_1 P_2 = P_4 P_3,
\end{eqnarray}
imply
\begin{eqnarray}
& &
s_1=0,\ s_2=0,\ -q_0 r_0=1 - \delta^2.
\label{eqleadingSolPart1}
\end{eqnarray}

The second set of relations
\begin{eqnarray}
& &
  \frac{a_1}{a_2}
=-\frac{a_4}{a_3}
=-\frac{b_2}{b_1}
= \frac{b_3}{b_4}
= \frac{i q_0}{1-\delta}
= \frac{1+\delta}{i r_0}
\end{eqnarray}
is solved as
\begin{eqnarray}
& & {\hskip -9.0truemm}
\left\lbrace
\begin{array}{ll}
\displaystyle{
a_1= a_{12} \lambda,\
b_2=-b_{12} \lambda,\
a_4= a_{34} \lambda,\
b_3= b_{34} \lambda,\
i q_0=(1-\delta) \lambda^2,\
}
\\
\displaystyle{
a_2= a_{12} \lambda^{-1},\
b_1= b_{12} \lambda^{-1},\
a_3=-a_{34} \lambda^{-1},\
b_4= b_{34} \lambda^{-1},\
i r_0=(1+\delta) \lambda^{-2},\
}
\end{array}
\right.
\label{eqLeadingCoeffsSol1}
\end{eqnarray}
in which $a_{12},a_{34},b_{12},b_{34},\lambda$ are nonzero
functions of $(z,t)$ to be determined.

Finally,
the two remaining equations
\begin{eqnarray}
& &
\left\lbrace
\begin{array}{ll}
\displaystyle{
a_{12} b_{12} - a_{34} b_{34} + \frac{I_0 C}{i      \gamma}
(1-\delta)(1-2 \delta)=0,
}
\\
\displaystyle{
a_{12} b_{12} - a_{34} b_{34} - \frac{I_0 C}{i \bar \gamma}
(1+\delta)(1+2 \delta)=0,
}\end{array}
\right.
\label{eqtworemaining}
\end{eqnarray}
first define $\delta$ as a root of the second degree equation
\begin{eqnarray}
& &
      \gamma (1+\delta)(1+2 \delta)
+\bar \gamma (1-\delta)(1-2 \delta)=0,
\label{eqleadingdelta1}
\end{eqnarray}
then put one constraint among $a_{12},b_{12},a_{34},b_{34}$.

The equation (\ref{eqleadingdelta1}) which defines $\delta$
only depends on the argument of $\gamma$,
\begin{eqnarray}
& &
(2 \cos g) \delta^2 +(3 i \sin g) \delta + \cos g=0,\
g=\arg \gamma.
\label{eqleadingdelta2}
\end{eqnarray}
When the photorefractive coupling constant $\gamma$ is
purely imaginary,
the exponent $\delta$ vanishes,
otherwise it can take two purely imaginary values
\begin{eqnarray}
& &
\delta=i \frac{- 3 \sin g \pm \sqrt{9 \sin^2 g + 8 \cos^2 g}}{4 \cos g}.
\end{eqnarray}

To conclude,
there generically exist two families (\ref{eqLeadingBehaviour})
of movable singularities,
defined by the equations
(\ref{eqLeadingPowersSol}),
(\ref{eqleadingSolPart1}),
(\ref{eqLeadingCoeffsSol1}),
(\ref{eqtworemaining}),
i.e.
\begin{eqnarray}
& &
\left\lbrace
\begin{array}{ll}
\displaystyle{
\delta=\hbox{ one the two roots of Eq.~}(\ref{eqleadingdelta2}),
}\\ \displaystyle{
p_1=p_4=P_2=P_3=-1+\delta,\
}\\ \displaystyle{
P_1=P_4=p_2=p_3=-1-\delta,\
}\\ \displaystyle{
a_1= \ \ N \lambda \ \ \ p_{12}\ \   \cosh \mu,\,\
b_2=-    N \lambda \ \ \ p_{12}^{-1} \cosh \mu,\,\
}\\ \displaystyle{
a_4= \ \ N \lambda \ \ \ p_{34}\ \   \sinh \mu,\,\
b_3= \ \ N \lambda \ \ \ p_{34}^{-1} \sinh \mu,\,\
}\\ \displaystyle{
a_2= \ \ N \lambda^{-1}  p_{12}\ \   \cosh \mu,\,\
b_1= \ \ N \lambda^{-1}  p_{12}^{-1} \cosh \mu,\,\
}\\ \displaystyle{
a_3=-    N \lambda^{-1}  p_{34}\ \   \sinh \mu,\,\
b_4= \ \ N \lambda^{-1}  p_{34}^{-1} \sinh \mu,\,\
}\\ \displaystyle{
i q_0=(1-\delta) \lambda^2,\
i r_0=(1+\delta) \lambda^{-2},\
}\\ \displaystyle{
N^2=-\frac{I_0 C}{i G} (1-\delta^2)(1-4 \delta^2),\
}\\ \displaystyle{
G=(1+\delta)(1+2 \delta) \gamma=-(1-\delta)(1-2 \delta)\bar \gamma.
}
\end{array}
\right.
\label{eqLeadingCoeffSol}
\end{eqnarray}
and they depend on four arbitrary complex functions
$\lambda,\mu,p_{12},p_{34}$ of $(z,t)$.

\subsection{Fuchs indices}
\label{sectionFuchs_indices}

The leading behaviour (\ref{eqLeadingBehaviour})
is the first term of a Laurent series
\begin{eqnarray}
& &
\chi \to 0:\
\left\lbrace
\begin{array}{ll}
\displaystyle{
    A_k = \chi^{p_k}  \sum_{j=0}^{+\infty} a_{k,j} \chi^j,\
     \de= \chi^\alpha \sum_{j=0}^{+\infty} q_j     \chi^j,\
}
\\
\displaystyle{
\barA_k = \chi^{P_k}  \sum_{j=0}^{+\infty} b_{k,j} \chi^j,\
  \dc   = \chi^\beta  \sum_{j=0}^{+\infty} r_j     \chi^j,\
}
\end{array}
\right.
\label{eqLaurent}
\end{eqnarray}
and the indices $j$ at which arbitrary coefficients enter this expansion
are computed as follows.
If one symbolically denotes $E(A_k,\barA_k,\de,\dc)=0$ the ten equations
(\ref{eq4wm}),
one builds the linearized system of (\ref{eq4wm}),
\begin{eqnarray}
& &
\lim_{\zeta \to 0}
\frac{E(A_k+\zeta C_k,\barA_k+\zeta \barC_k,
          \de+\zeta D,  \dc  +\zeta \barD)
-E(A_k,\barA_k,\de,\dc)}{\zeta}
=0,
\nonumber
\end{eqnarray}
i.e.
\begin{eqnarray}
& &
\left\lbrace
\begin{array}{ll}
\displaystyle{
\partial_z C_1 + i \de C_2 + i A_2 D=0,\ \dots
}
\\
\displaystyle{
\left(\partial_t +\frac{1}{\tau}\right) D - \frac{\gamma}{I_0}
\left(\barA_2 C_1 + A_1 \barC_2 + \barA_3 C_4 + A_4 \barC_3  \right)=0.
}
\end{array}
\right.
\label{eqLinearized2}
\end{eqnarray}
At the point (\ref{eqLeadingBehaviour}),
this ten-dimensional linear system displays near $\chi=0$
a singularity which has the Fuchsian type \cite[Chap.~15]{Ince},
therefore it admits a solution
\begin{eqnarray}
& &
\chi \to 0:\
\left\lbrace
\begin{array}{ll}
\displaystyle{
C_k \sim c_k \chi^{p_k+j},\ \barC_k \sim d_k \chi^{P_k+j},\
k=1,2,3,4,\
}
\\
\displaystyle{
D \sim s_0 \chi^{\alpha+j},\ \barD \sim t_0 \chi^{\beta+j}.
}
\end{array}
\right.
\end{eqnarray}
The condition that this solution be nonidentically zero
results in the vanishing of a tenth order determinant
whose roots $j$, called Fuchs indices,
take the ten values
\begin{eqnarray}
& &
j=-1,0,0,0,0,2,2,2,\frac{5 \pm \sqrt{1+48 \delta^2}}{2}.
\label{eqindices}
\end{eqnarray}
In the two-wave mixing case,
the four roots $0,0,2,2$ are to be removed from this list.
The necessary condition, required by the Painlev\'e test,
that all indices be integer is violated when $\delta$ is nonzero,
therefore the \FWM\ model fails the test in this case.
This failure is quite similar to what occurs in the
complex cubic Ginzburg-Landau equation \cite{CT1989},
and conclusions similar to those of Ref.~\cite{TF}
can probably be drawn.

Let us remark that the Fuchs index $0$ has multiplicity four,
in agreement with the number of
arbitrary functions involved in the leading behaviour,
see Eq.~(\ref{eqLeadingCoeffSol}).

\subsection{Conditions at the Fuchs indices}
\label{sectionLeading1}

The coefficients $a_{k,j},b_{k,j},q_j,r_j$ of the Laurent series
(\ref{eqLaurent}) are computed for $j \ge 1$ by solving a linear system.
Whenever $j$ reaches an integer Fuchs index,
an obstruction may occur,
resulting in the introduction of a movable logarithmic branching,
and some necessary conditions need to be satisfied in order to avoid it.
In the generic case $\delta \not=0$,
these obstructions may only arise at $j=2$, value of a triple Fuchs index.
In the nongeneric case $\delta=0$,
in addition to the quadruple index $j=2$,
one must also check the simple index $j=3$.
The results are as follows.

At $j=2$, whatever be $\delta$, a movable logarithm exists,
unless the following necessary condition is satisfied,
\begin{eqnarray}
& &
Q_2 \equiv a \left(C_t + C C_z - 2 a C \right)=0,\ a=\frac{1}{\tau}.
\label{eqQ2cond}
\end{eqnarray}
Since $a=0$ is excluded,
the second factor defines a precise dependence of $C$ on $(z,t)$,
hence a similar dependence for $S,\chi,\varphi$
and ultimately for the five complex amplitudes \textit{via}
their Laurent expansion (\ref{eqLaurent}).
Let us find this dependence explicitly.

The general solution of (\ref{eqQ2cond}) is
provided by the method of characteristics
and is defined implicitly by the relation
\begin{eqnarray}
& &
2 a z= C +F(e^{-2 a t} C),\
\label{eqQ2C}
\end{eqnarray}
in which $F$ is an arbitrary function of one variable.

The invariant $S$ is then defined by the first order linear PDE
(\ref{eqCrossXT}),
which reads
\begin{eqnarray}
& &
S(x,t)=\Sigma(C,t),\
D=1 + e^{-2 a t} F',\
\nonumber
\\
& &
e^{2 a t} \left(\Sigma_t + 2 a C \Sigma_C + 4 a \frac{\Sigma}{D}\right)
+ 24 a^3 e^{-6 a t} \frac{{F''}^2}{D^5}
-  8 a^3 e^{-4 a t} \frac{F'''}{D^4}=0,
\end{eqnarray}
and admits the general solution
\begin{eqnarray}
& &
S=
c_0    e^{-4 a t} D^{-2}
-4 a^2 e^{-6 a t} D^{-3} F'''
+6 a^2 e^{-8 a t} D^{-4} {F''}^2,
\label{eqQ2S}
\end{eqnarray}
in which $c_0$ is an arbitrary constant.

We have not been able to prove whether the function $F$
is a gauge which can be arbitrarily chosen (e.g. $F=0$)
or whether it is essential.
If it is essential,
there could exist much more intricate solutions than those which we
now outline.

When $F$ is arbitrary,
we have not yet succeeded to integrate the system
(\ref{eqChixt})
for $\chi(z,t)$,
hence to compute the induced dependence of
$A_j,\de$ on $(z,t)$.
However, if one restricts to a constant value for $F$,
which can then be chosen equal to zero,
the equation (\ref{eqQ2C}) for $\varphi$ integrates as
\begin{eqnarray}
& &
F=0,\
\varphi=\Phi(\xi),\
\xi=\sqrt{z}e^{-a t},
\label{eqphival}
\end{eqnarray}
in which $\Phi$ is an arbitrary function,
and it is straightforward to check that the Laurent expansion
(\ref{eqLaurent}) defines the reduction $(z,t) \to \xi=\sqrt{z} e^{-a t}$.
As opposed to the stationary reduction $\partial_t=0$,
this
 reduction is noncharacteristic
(i.e.~it does not lower the differential order)
it is defined with a reduced variable of the factorized type
$\xi=f(z) g(t)$,
\begin{eqnarray}
& &
\left\lbrace
\begin{array}{ll}
\displaystyle{
\xi=z^{1/2} e^{-a t},\ a=1/ \tau,
}\\ \displaystyle{
\de(z,t)=e^{-a t-i \omega t} z^{-1/2} \de_{r}(\xi),\
\dc(z,t)=e^{-a t+i \omega t} z^{-1/2} \dc_{r}(\xi),\
}\\ \displaystyle{
    A_j(z,t)=e^{-a t-i \omega t/2}     A_{j,r}(\xi),\
\barA_j(z,t)=e^{-a t+i \omega t/2} \barA_{j,r}(\xi),\ j=1,4,
}\\ \displaystyle{
    A_j(z,t)=e^{-a t+i \omega t/2}     A_{j,r}(\xi),\
\barA_j(z,t)=e^{-a t-i \omega t/2} \barA_{j,r}(\xi),\ j=2,3, }\\
\displaystyle{ \frac{\D}{\D \xi}     A_{1,r}=-2 i \de_r
A_{2,r},\ \frac{\D}{\D \xi} \barA_{2,r}= 2 i \de_r \barA_{1,r},\
}\\ \displaystyle{
\frac{\D}{\D \xi} \barA_{3,r}=-2 i \de_r \barA_{4,r},\
\frac{\D}{\D \xi}     A_{4,r}= 2 i \de_r     A_{3,r},\ }\\
\displaystyle{ \frac{\D}{\D \xi} \de_r +i \omega \tau \xi^{-1} \de_r
=-\frac{\gamma}{I_0} \left(A_{1,r} \barA_{2,r} + \barA_{3,r}
A_{4,r} \right), }
\end{array}
\right.
\label{eq4wrReducxi}
\end{eqnarray}
and it depends on the additional real parameter $\omega$.

To summarize the information obtained from the Fuchs index $2$,
provided $F$ can be arbitrarily chosen,
the four-wave mixing model (\ref{eq4wm})
(as well as the two-wave mixing model) admits no
single valued dynamical solution other than the possible solutions
of the reduction $(z,t) \to \xi=\sqrt{z} e^{-a t}$.

When $\delta\not=0$,
this ends the Painlev\'e test.
When $\delta=0$,
the Fuchs index $3$
is found to be free of movable logarithms.

Finally,
the information provided by the Painlev\'e test is the following.
\begin{enumerate}
\item
Whatever be $\gamma$,
and under the mild restriction that $F$ can be arbitrarily chosen,
no single valued dynamical solution of (\ref{eq4wm}) exists other than
the possible solutions of the reduction $(z,t) \to \xi=\sqrt{z}e^{-a t}$.

\item
For this reduction, two cases must be distinguished.
When $\Re(\gamma) \not=0$,
the ten-dimensional ODE system (\ref{eq4wrReducxi})
possesses at most an eight-parameter single valued solution.
When $\Re(\gamma)=0$,
the system (\ref{eq4wrReducxi}) passes the Painlev\'e test
therefore it may admit a ten-parameter single valued solution,
which then would be its general solution.
Finding a Lax pair in this case would considerably help
to perform the explicit integration.

\end{enumerate}

Similar conclusions apply to the six-dimensional two-wave mixing:
the only possibility for a single valued solution
is the reduction (\ref{eq4wrReducxi}) (with $A_3=A_4=0$),
this solution depending on
four movable constants for $\Re(\gamma) \not=0$,
and
six movable constants for $\Re(\gamma)=0$.

\section{Conclusion}

The present study proves, like for the complex Ginzburg-Landau equation,
that physically relevant analytic solutions quite certainly exist
for the four-wave mixing model.
The present counting of the possible arbitrary constants in the solutions
displays, as expected, the crucial role of the photorefractive complex constant.
Explicit solutions based on the present study will be presented elsewhere.

The results can be used for predicting new phenomena in optical self-diffraction
of waves in
photorefractive media which use the non-local response as well as for
optimization of optical
dynamic holographic settings.
Among these applications let us quote:
 (i) the formation of a localized grating to increase optical information density;
(ii) the methods of all-optical control of
output wave characteristics versus input beam intensities and phases;
(iii)
the optimization of the parameters of optical phase-conjugation;
(iv) the use of the new holographic topographic
technique to material parameter characterization.
Just like similar processes
of nonlinear self-action of waves arise in models of optical networks,
optical information processing,
quantum information processing, interacted neural chains,
then various other problems of nonlinear
wave interaction can become the subject of further independent research.

\section*{Acknowledgments}

We warmly acknowledge the financial support of the
Max Planck Institut f\"ur Physik komplexer Systeme,
and RC thanks the WASCOM organizers for invitation.


\vfill\eject
\end{document}